\documentclass[aps,prb,twocolumn,superscriptaddress,showpacs,10pt]{revtex4}
\usepackage{latexsym}
\usepackage{graphicx}
\usepackage{graphicx}% Include figure files
\usepackage{dcolumn} % Align table columns on decimal point
\usepackage{bm}
\usepackage{epsfig}
\begin{document} 

\title{Ab initio study of the influence of nanoscale doping inhomogeneities in the phase separated state of La$_{1-x}$Ca$_{x}$MnO$_3$}

\author{A. Pi\~neiro}
\affiliation{Departamento de F\'{i}sica Aplicada,
  Universidad de Santiago de Compostela, E-15782 Campus Sur s/n,
  Santiago de Compostela, Spain}
\affiliation{Instituto de Investigaciones Tecnol\'{o}gicas,
  Universidad de Santiago de Compostela, E-15782 Campus Sur s/n,
  Santiago de Compostela, Spain}
\author{V. Pardo}
\email{victor.pardo@usc.es}
\affiliation{Departamento de F\'{i}sica Aplicada,
  Universidad de Santiago de Compostela, E-15782 Campus Sur s/n,
  Santiago de Compostela, Spain}
\affiliation{Instituto de Investigaciones Tecnol\'{o}gicas,
  Universidad de Santiago de Compostela, E-15782 Campus Sur s/n,
  Santiago de Compostela, Spain}
\author{D. Baldomir}
\affiliation{Departamento de F\'{i}sica Aplicada,
  Universidad de Santiago de Compostela, E-15782 Campus Sur s/n,
  Santiago de Compostela, Spain}
\affiliation{Instituto de Investigaciones Tecnol\'{o}gicas,
  Universidad de Santiago de Compostela, E-15782 Campus Sur s/n,
  Santiago de Compostela, Spain}
\author{A. Rodr\'{i}guez}
\affiliation{CESGA,
  E-15782 Avda. de Vigo s/n,
  Santiago de Compostela, Spain}
\author{R. Cort\'{e}s-Gil}
\affiliation{Departamento de Qu\'{i}mica Inorg\'{a}nica, Facultad de C.C. Qu\'{i}micas,
  Universidad Complutense de Madrid, 28040,
  Madrid, Spain}
\author{A. G\'{o}mez}
\affiliation{CESGA,
  E-15782 Avda. de Vigo s/n,
  Santiago de Compostela, Spain}
\author{J.E. Arias}
\affiliation{Instituto de Investigaciones Tecnol\'{o}gicas,
  Universidad de Santiago de Compostela, E-15782 Campus Sur s/n,
  Santiago de Compostela, Spain}
  
\date{\today}

\begin{abstract}
The chemical influence in the phase separation phenomenon that occurs in perovskite manganites is discussed by means of ab initio calculations. Supercells have been used to simulate a phase separated state, that occurs at Ca concentrations close to the localized to itinerant crossover. We have first considered a model with two types of magnetic ordering coexisting within the same compound. This is not stable. However, a non-isotropic distribution of chemical dopants is found to be the ground state. This leads to regions in the system with different effective concentrations, that would always accompany the magnetic phase separation at the same nanometric scale, with hole-rich regions being more ferromagnetic in character and hole-poor regions being in the antiferromagnetic region of the phase diagram, as long as the system is close to a phase crossover.
\end{abstract}

\maketitle

%\newpage

\section{Introduction}

Perovskite manganites belong to a class of materials that have become increasingly important in the last few years, particularly because of the large variety of phenomena they exhibit, closely tied to the strongly correlated nature of their electronic structure. The series of manganites A$_{1-x}$B$_{x}$MnO$_3$ (A$^{3+}$ being a lanthanide and B$^{2+}$ an alkaline earth) is one of the most intensively studied families.\cite{pr_magnanites} The interest in these materials grew significantly after the discovery of colossal magnetoresistance (CMR), a large negative magneto-resistance taking place under an applied magnetic field,\cite{rao} but also considerable research has focused on their unusual properties such as phase separation,\cite{shenoy} charge and orbital ordering\cite{dagotto,wu} and also for their potential applications in thin-film devices,\cite{bowen,tsymbal,eerenstein} derived of their peculiar electronic structure properties.

The case of A = La$^{3+}$ and B = Ca$^{2+}$ has become a prototypical example of manganite compounds,\cite{zhang,herrero} showing a very rich phase diagram.\cite{schiffer,phase} The region in the phase diagram with concentrations of about 18\% and 50\% Ca$^{2+}$ have been thoroughly studied, because  samples exhibit phase separation, intrinsic to the proximity to the magnetic/electronic phase transition. In that part of the phase diagram, La$_{1-x}$Ca$_{x}$MnO$_3$ presents mixed Mn$^{3+}$/Mn$^{4+}$ valencies and a strong electron-phonon coupling, the magnetic coupling being a modified version of the double exchange interaction to take into account the coupling between the electronic states and vibrational modes (vibrons). The mechanism for ferromagnetic (FM) ordering in manganites exhibiting CMR has been recently considered,\cite{prl_marta} suggesting a breakdown of the adiabatic approximation for the itinerant electrons. These results are related with dynamic phase segregation into hole-rich itinerant-electron regions and hole-poor localized-electron regions, leading to suppression of the conventional double-exchange mechanism.

Several theoretical studies have been performed in these manganite perovskite systems using different techniques. The electronic structure of La$_{1-x}$Ca$_x$MnO$_3$ was calculated using the LAPW (Linearized Augmented Planewave) method.\cite{lapw1,lapw2} Transport properties were studied using the virtual crystal approximation, neglecting disorder effects in the lattice.\cite{lacamno_vca} The formation of a pseudogap follows from numerical calculations based on the dynamical mean field theory, indicating the existence of a charge-nonuniform ground state.\cite{lacamno_dmft} An extensive study of the physics of manganites, including the structure we are studying here, was carried out by Salomon et al.\cite{salomon} Unfortunately, an ab initio description of the phase separation in manganites has not been given in the literature.

It is now widely accepted that some sort of electronic phase segregation is at the origin of the CMR effect in manganites. Moreo \textit{et al.}\cite{moreo} proposed a scenario where the coexistence of two phases with different charge density is unstable against breakup into nanometer size clusters due to the Coulomb force. Two different kinds of electronic phase segregation have been proposed to exist in manganites:\cite{book_dagotto} equal density vs. different density electronic phase segregation at the micro- and nano-scale, respectively. Experimentally, magnetic phase separation has been observed in the form of nanoscale phases with different magnetic properties,\cite{hennion} i.e. regions in the crystal with nanometric size that present a different magnetic ordering. Imaging techniques\cite{prl_new} have recently shown the close relationship between nanoscale phase separated regions of an electronic/magnetic origin and CMR. According to those studies, these regions, on the several nm range, have not chemical origin. A different view comes from an analysis of the influence of the dopant distribution in the superconductivity and phase separation that has recently appeared\cite{yeoh} studying the chemical inhomogeneities at the nanoscale in (Ba$_{1-x}$K$_x$)Fe$_2$As$_2$ superconducting pnictides.

The fundamental problem is to give a correct description of how the holes are distributed in the system. It has been known for long that even the simplest models\cite{emery} predict a non-uniform density distribution of dopants, which are unstable against segregation into hole-rich and hole-poor regions. To gain insight into this problem, in this paper, we have taken into account a structure with Ca$^{2+}$ doping levels close to $x\sim 0.2$ (exactly $x = 0.1875$) of La$_{1-x}$Ca$_{x}$MnO$_3$ from a computational point of view via ``ab initio" calculations. Typically, this is not done, and models are utilized instead, due to the large supercells that are required, but even the crudest supercells can help us, as we will see, to understand the electronic and chemical origins of phase separation. It is worth recalling that the above mentioned doping level is experimentally placed around the critical boundary in which the FM insulating to FM metallic transition occurs at low temperatures with respect to doping.\cite{schiffer,phase} We have constructed several superstructures in different magnetic configurations to explore the magnetic origin of phase separation. Since our calculations yield a homogeneous FM ground state, we will analyze the influence of the Ca$^{2+}$ dopant distribution in the lattice in this magnetic ground state. Varying the Ca$^{2+}$ positions from a random distribution to a slightly correlated one, we can analyze the influence of dopant distribution in the electronic structure and the physical properties of the compound by means of ``ab initio" calculations. 

For the sake of completeness, we will make a similar analysis in a very different Ca$^{2+}$ concentration ($x =0.375$), in which the ground state is FM metallic becoming paramagnetic insulating at high temperatures. The idea of that calculation is to study a concentration away from the phase boundaries at $x = 0.1875$ and also from $x = 0.5$, where the ground state becomes an insulating antiferromagnetic (AF) and charge ordered (CO) state. At those two hole densities there is a strong tendency towards phase separation caused by intrinsic inhomogeneities\cite{pr_magnanites} but $x = 0.375$ (unlike $x = 0.1875$ and $0.5$) is not close to the boundary between different ground states in the La$_{1-x}$Ca$_{x}$MnO$_3$ solid solution.\cite{schiffer,phase} These analyses will help us discuss the consequences of having chemical inhomogeneities on the nanoscale in these phase separated compounds.

\section{Computational details}

Electronic structure calculations were  performed within the density functional theory framework \cite{dft} using {\sc WIEN2k} software,\cite{wien, wien2k} which utilizes an augmented plane wave plus local orbitals (APW+lo) \cite{sjo} method to solve the Kohn-Sham equations. This method uses an all-electron, full-potential scheme that makes no shape approximation to the potential or the electron density.

For modeling the correlated behavior of the d electrons of the system, we have included strong correlation effects by means of the LDA+U scheme,\cite{sic} controlled by an effective U (U$_{eff}$=U-J), being U the on-site Coulomb repulsion and J the on-site exchange constant (taken as J=0, as is common practice in literature).\cite{ylvisaker} Values of U$_{eff}$: 2.7, 4.0 and 5.5 eV were used in the calculations. These values are on the range that have been used in the past for various Mn oxides.\cite{u_mn1,u_mn2} The structural minimization was carried out using the GGA$-$PBE scheme\cite{gga} (generalized gradient approximation in the Perdew-Burke-Ernzerhof scheme) minimizing the forces on the atoms and the total energy of the system. The parameters of our calculations \cite{singh} depend on the type of calculation but for any of them we converged with respect to the \textit{k}-mesh and to $R_{mt}K_{max}$, up to 75 $k$ points (15 in the irreducible Brillouin zone, in a 3 $\times$ 3 $\times$ 5 sampling of the full Brillouin zone for a supercell N $\times$ N $\times$ N perovskite unit cells) and up to $R_{mt}K_{max}=6.0$. Muffin tin radii chosen were the following: 2.43 a.u. for La, 2.29 a.u. for Ca, 1.96 a.u. for Mn and 1.73 a.u. for O.

\section{Results and discussion}

\subsection{Magnetic study}

\begin{figure}
\includegraphics[width=\columnwidth,draft=false]{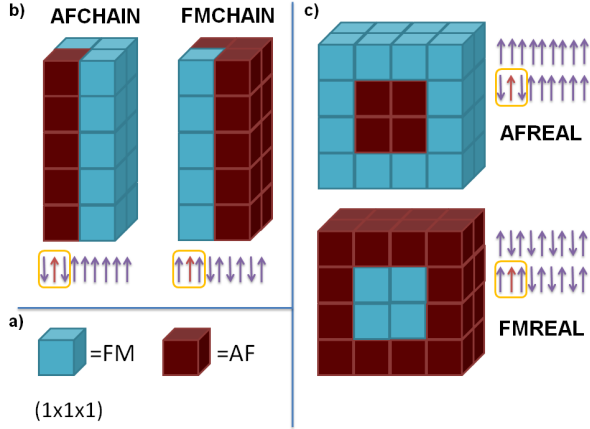}
\caption{(Color online) Schematic cases of the phase separated structures studied. On the bottom left we see a representation of the unit cell of La$_{0.8125}$Ca$_{0.1875}$MnO$_{3}$ (1$\times$1$\times$1) in the ferromagnetic (FM, blue) configuration and in the antiferromagnetic (AF, red) configuration (a). On the upper left (b) we represent the (2$\times$2$\times$1) cases, one dimensional AF chains surrounded by a FM guide and vice versa. On the right (c) we represent both realistic cases of phase separation (2$\times$2$\times$2): an AF nanometric region completely surrounded by a FM region (upper) and vice versa (lower). In this latter case, a 4$\times$4$\times$2 is depicted for clarity.}\label{cases_magnetic}
\end{figure}

The goal of the first part of our work is to simulate a magnetically phase separated state by embedding a FM (AF) phase into an AF (FM) matrix. For this reason, we have set up several superstructures based on the unit cell La$_{0.8125}$Ca$_{0.1875}$MnO$_{3}$ (at a concentration close to a metal-insulator transition in the phase diagram), using a 2$\times$1$\times$2 and a 2$\times$2$\times$2 supercell (in perovskite unit cell units), as shown in Fig. \ref{cases_magnetic}. The larger and initially more realistic cases (2$\times$2$\times$2) in Fig. \ref{cases_magnetic} (c), serve as a model of the experimentally observed situation of a magnetic phase (antiferromagnetic (AFREAL) in the upper figure, ferromagnetic (FMREAL) in the lower one) completely surrounded by a magnetically different phase (FM and AF, respectively). Figure \ref{cases_magnetic} (b) shows the cases of an AF chain surrounded by a FM guide (AFCHAIN) and vice versa (FMCHAIN). We have also calculated the system in the simple ferromagnetic (FM) and G-type AF cases (Fig. \ref{cases_magnetic} (a)).

\begin{figure}
\includegraphics[width=\columnwidth,draft=false]{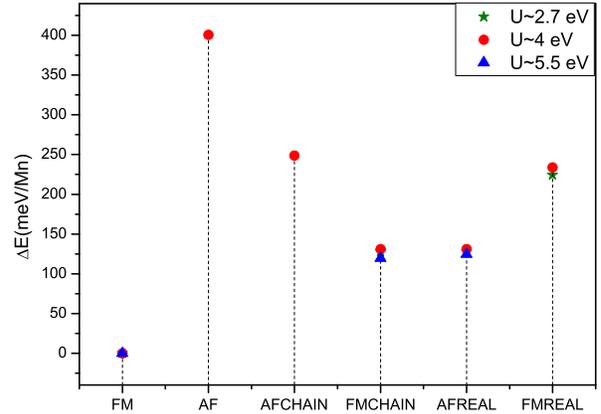}
\caption{(Color online) Calculated energies of the constructed structures at several values of U within the LDA+U approximation. We take the more stable calculated case (FM case) as the zero energy state. All energies are expressed in meV/Mn.}\label{energies_away}
\end{figure}

We have calculated all these phase separated superstructures and compared their relative stability (total energy calculations) with respect to an entirely FM structure and also an entirely AF structure (G-type). Results are summarized in Fig. \ref{energies_away}. We can see that the purely FM structure is the most stable one in every case, the stabilization energy ranging from 130 meV/Mn compared to the FMCHAIN and AFREAL structures to 400 meV/Mn with respect to the AF structure. Calculations predict that these magnetically phase separated states are not stable. The picture is the same for all values of U we have calculated. From these calculations, we can conclude that the FM configuration at the concentration $x_0 = 0.1875$ is the most stable, in agreement with experimental observations at that concentration.\cite{schiffer,phase} These results suggest that the coexistence of two magnetic phases in the compound require some other changes in the crystallographic and electronic structure for it to be stable.

%Observing all these results, it becomes clear that magnetism alone cannot drive the system to a phase separated state with electronically segregated phases. Other mechanisms might be playing a role.

\subsection{Chemical study}

Holes in the e$_g$ Mn levels can be introduced by the substitution of La$^{3+}$ by Ca$^{2+}$ over the entire compositional range in the La$_{1-x}$Ca$_{x}$MnO$_3$ solid solution. Additionally, hole doping can be created by means of cationic vacancies. 
La$_{0.8125}$Ca$_{0.1875}$MnO$_{3}$ is a distorted perovskite with a pseudocubic lattice parameter (a$_c$) of 3.89 \AA. The tilting of the MnO$_6$ octahedra results in an orthorhombic symmetry (space group Pnma, no. 62) with lattice parameters $a_o \approx c_o \approx \sqrt{2} a_c$ and $b_o \approx 2 a_c$ for this composition. The ability to oxidize La$_{1-x}$Ca$_{x}$MnO$_3$ materials with $x \leq 0.2$ leads to important structural, electronic and magnetic variations related to the presence of cationic vacancies.\cite{herrero,alonso_prb,alonso_chemmat} Thus, differences between non-stoichiometric and stoichiometric samples should be always considered.

Stoichiometric La$_{0.8125}$Ca$_{0.1875}$MnO$_3$ compound, i.e., with no cationic (and anionic) vacancies is considered in the present study. We initially have assumed that the La$^{3+}$ and Ca$^{2+}$ atoms (the dopants) are distributed in a perfectly homogeneous manner throughout the crystal. But, experimentally, this might not be the case; nano-sized chemical inhomogeneities in the structure can happen and will not be detected below the few nm limit,\cite{prl_new} but they could have a key role in the properties of a phase separated state.

\begin{figure}
\includegraphics[width=\columnwidth,draft=false]{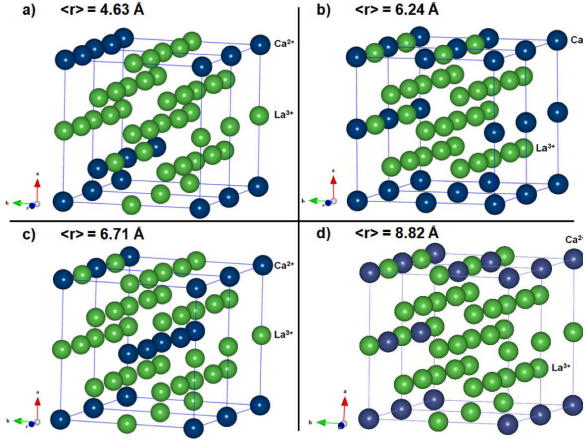}
\caption{(Color online) Schematic representation of Ca$^{2+}$ atoms distribution in the lattice of La$_{0.8125}$Ca$_{0.1875}$MnO$_{3}$. $<$r$>$ is the average distance between Ca$^{2+}$ atoms in the lattice. These figures show the four possible atomic ordering of Ca$^{2+}$ atoms in the lattice (because of its symmetry) and its $<$r$>$ from the most inhomogeneous case distribution (a) to the homogeneous case distribution (d).}\label{schematic_r}
\end{figure}

The next step of the work is to study if such dopant inhomogeneities can occur in the nanoscale. To do this, we have varied the atomic positions of the Ca$^{2+}$ atoms to distribute them in various manners throughout the crystallographic unit cell. In the La$_{0.8125}$Ca$_{0.1875}$MnO$_{3}$ structure we can change the distribution of the Ca$^{2+}$ cations to create four possible inequivalent chemical configurations. These can be seen in Fig. \ref{schematic_r}. We can define the quantity $<$r$>$ as the average distance between Ca$^{2+}$ atoms in the lattice, and use it as a measure of how homogeneously these dopants are distributed:

\begin{equation}
<r>=\sum^{N}_{i=1}\frac{d_{i}}{N}
\end{equation}

where $d_{i}$ is the distance between nearest Ca$^{2+}$ atoms and $N$ is the number of nearest Ca$^{2+}$ atoms. 

In principle, inserting chemical inhomogeneities into the lattice will generate uncompensated forces and increase the elastic energy in the structure, but we need to calculate all the energetic terms involved to obtain the most stable configuration. In our case, we will do that using ab initio calculations.

\begin{figure}
\includegraphics[width=\columnwidth,draft=false]{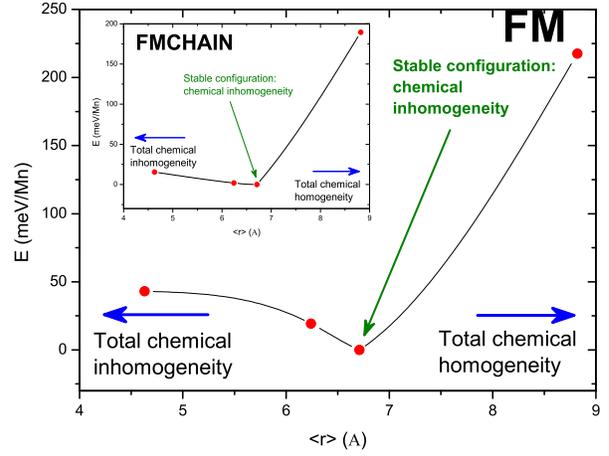}
\caption{(Color online) Energies of the calculated structures for La$_{0.8125}$Ca$_{0.1875}$MnO$_{3}$ (y axis) in the FM configuration. The energies are normalized to the lowest value of the four possible Ca$^{2+}$ chemical distribution configurations. On the x axis we write the values of $<$r$>$ for each structure (see text), from the larger one on the right (total homogeneity structure corresponds with Fig. \ref{schematic_r} d) to the shorter one on the left (total inhomogeneity structure corresponds with Fig. \ref{schematic_r} a). The inset shows an analogous study of the second magnetic stable configuration calculated in section A of this paper, which is the FMCHAIN configuration.}\label{la08_r}
\end{figure}

We have constructed these four possible cases using a FM ordering, the most stable magnetic structure we explored previously. We have performed a full lattice relaxation for each of the cases, and then we have calculated their total energies. The results are presented in Fig. \ref{la08_r}. We can observe that the most stable structure corresponds to an intermediate case between the most homogeneous (Fig. \ref{schematic_r}d) and the most inhomogeneous (Fig. \ref{schematic_r}a) case, which is the one used for the calculations in previous sections. Taking the ground state structure as the $<$r$>$= 6.71 \AA, we found this one is more stable by about 19 meV/Mn with respect to the second most stable configuration, and by 219 meV/Mn to the most unstable configuration (largest chemical homogeneity attainable at this concentration). This means that a non-homogeneous dopant distribution occurs in the material that would lead to a nanoscale distribution of effective concentrations. Analogous results in terms of dopant distribution energetics can be found in other magnetic configurations (hence, it is not an issue of the FM solution).

\begin{figure}
\includegraphics[width=\columnwidth,draft=false]{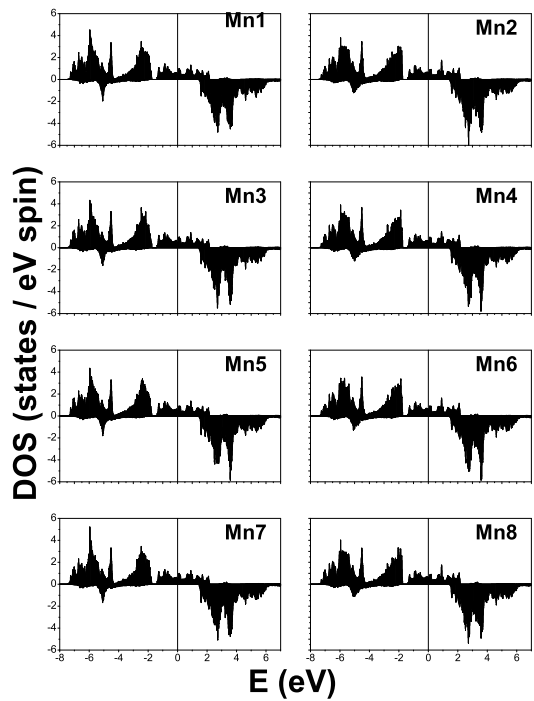}
\caption{(Color online) Majority spin (positive values) and minority spin (negative values) partial density of states (DOS) plots of the eight inequivalent Mn atoms in the ground state (both chemically and magnetically) configuration. Fermi energy is represented by a vertical line at E=0 eV. }\label{DOS_mn}
\end{figure}

The most stable $<r>$ value is larger than one cubic perovskite lattice parameter. Several perovskite subcells will be involved in forming this ground state Ca$^{2+}$ distribution. Moreover, variations of the atomic position of the various Ca$^{2+}$ cations will imply different environments for each of the Mn atoms in the unit cell. Thus, the existence of areas with such a typical size suggests the magnetic and electronic characteristics differ inside and outside the clusters leading to an electronically inhomogeneous state on a nanometric scale in La$_{0.8125}$Ca$_{0.1875}$MnO$_{3}$. Zones where Ca$^{2+}$ atoms are grouped lead to a local concentration displacement towards the hole rich region (Ca$^{2+}$-rich) in the phase diagram of La$_{1-x}$Ca$_{x}$MnO$_{3}$. Instead, zones where the density of Ca$^{2+}$ atoms are lower, we have a La$^{3+}$-rich (electron-rich) zone, being a local concentration displacement to the AF region of the phase diagram. This scenario is consistent with the stable FM order\cite{alonso_prb} near Mn$^{4+}$, localized near Ca$^{2+}$-rich regions, in agreement with the hole distribution described by Alonso \textit{et al.} as the hole attractor model.\cite{alonso_chemmat} Our calculations show that the existing chemical inhomogeneities can be a coexisting factor that works together with the electronic/magnetic phase separation.

We have calculated the electronic structure of the material in the ground state (both in terms of magnetic ordering and chemical configuration). The values obtained for the total magnetic moment in this structure are of 3.8125  $\mu_B$ per Mn atom, as expected. This corresponds to an average Mn valence of +3.1875. We can also analyze the magnetic moments inside the muffin-tin spheres for each of the Mn atoms. These are summarized in Table \ref{mn_mmi}. Their magnetic moments present differences on the order of 1 $\%$ at most. Also the total $d$ charge inside the Mn muffin-tin spheres varies by only 0.015 electrons. This all would be translated in changes in effective Ca concentration $x$ of only 0.01. This can be seen in the largely homogeneous electronic structure observed in the partial density of states of the inequivalent Mn atoms in the ground state of the compound (Fig. \ref{DOS_mn}). In a metallic solution, the way the different local concentrations correlate with an electronic phase separation in the system is not obvious, but we observe that the chemical inhomogeneities relate to a different electron count in each Mn atom.

\begin{table}[h!]
  \caption{Values of the atomic magnetic moment in units of $\mu_{B}$ inside the muffin-tin sphere for the ground state configuration. Values of total magnetic moment are 3.8125 $\mu_{B}$ per Mn atom, with an average valence Mn$^{3.1875+}$, the value expected from the electron count in a fully polarized half-metallic ferromagnet.}
  \begin{center}
  \begin{tabular}{cccccc}
    \hline
    \hline
    &  & Magnetic & & & Magnetic \\
    &  & Moments & & & Moments \\
    \hline
    Mn1 & & 3.56 & Mn5 & & 3.56 \\
    Mn2 & & 3.52 & Mn6 & & 3.52 \\
    Mn3 & & 3.56 & Mn7 & & 3.56 \\
    Mn4 & & 3.52 & Mn8 & & 3.52 \\
    \hline
  \end{tabular}
  \end{center}
  \label{mn_mmi}
\end{table}

\begin{figure}
\includegraphics[width=\columnwidth,draft=false]{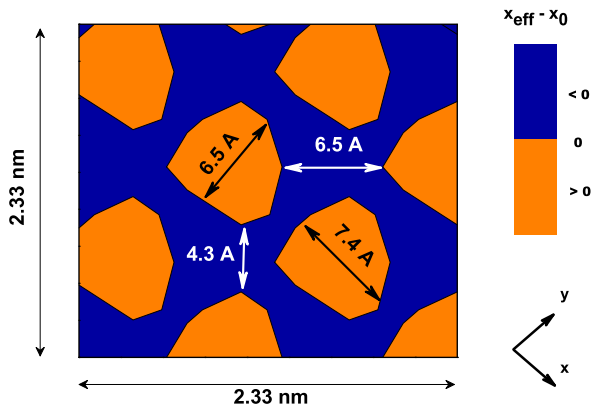}
\caption{(Color online) Difference between local concentrations and global concentration of La$_{0.8125}$Ca$_{0.1875}$MnO$_{3}$.  Orange (lighter) colors indicate a hole rich region on the phase diagram of La$_{1-x}$Ca$_{x}$MnO$_{3}$,\cite{phase} leading to a concentration where FM is the ground state. Blue (darker) colors indicate an electron rich region on the phase diagram (where the ground state is AF). The diagram shows zones of about 7.0 \AA\ in typical size with Ca concentration $x_{eff}$ $>$ $x_0$ and other zones of about 5.5 \AA\ with $x_{eff}$ $<$ $x_0$ ( $x_0 = 0.1875$ ).}\label{local_doped}
\end{figure}

The typical sizes of these hole/electron rich zones that turn out from our calculations for the ground state chemical distribution can be seen in Fig. \ref{local_doped}. We have calculated the effective $x$ for each inequivalent Mn atom in the supercell according to its nearest neighbours (up to fourth neighbours), assuming the usual Ca$^{2+}$/La$^{3+}$ valencies, for this most stable chemical configuration. We observe there a  $6.5 - 7.4$ \AA\ diameter zone where the effective doping of La$_{1-x}$Ca$_{x}$MnO$_{3}$ has more Ca$^{2+}$ than the real macroscopic average concentration ($x=0.1875$) and a $4.3 - 6.5$ \AA\ diameter zone where the effective doping is displaced to the La$^{3+}$-rich zone in the phase diagram. These calculated cluster sizes are comparable, but somewhat lower than those reported in literature\cite{cortesgil_chemmat} concerning the Ca-rich region in the La$_{1-x}$Ca$_{x}$MnO$_{3}$ solid solution. They show the importance of chemical ordering at the nanoscale when the system is close to a metal-insulator transition. The presence of chemical inhomogeneities on a similar nanometric scale than the experimentally observed electronic/magnetic phase separation could act as a precursor for the phase separated scenario to develop in the system. The largest deviation around the $x_{0}=$0.1875 in terms of Ca concentration ($x$) is about 0.04 (slightly larger than the equivalent deviation in the magnetic moments of Mn atoms, which was calculated to be about 0.01). Chemical variations are stronger than the electronic variations. Changes in the local doping of each Mn will vary the effective Ca concentration from $0.167<x<0.208$, which are enough to move the system to the other side of the magnetic/electronic phase transition, that occurs at about $x = 0.18$. In fact, different nanosize domains have been experimentally detected around this compositional range: FM clusters of 16 \AA\ for $x = 0.17$ whereas an approximate size of 8 \AA\ has been observed for $x = 0.2$.\cite{hennion} This last result confirms the existence of FM clusters confined in about two perovskite subcells, in a similar scale than the chemical inhomogeneities we find in our calculations for $x = 0.1875$. The zones with a lower Ca$^{2+}$ concentration will have an AF ground state and the zones with higher values of $x$ will be in their FM ground state, according to the phase diagram of the system.

\begin{figure}
\includegraphics[width=\columnwidth,draft=false]{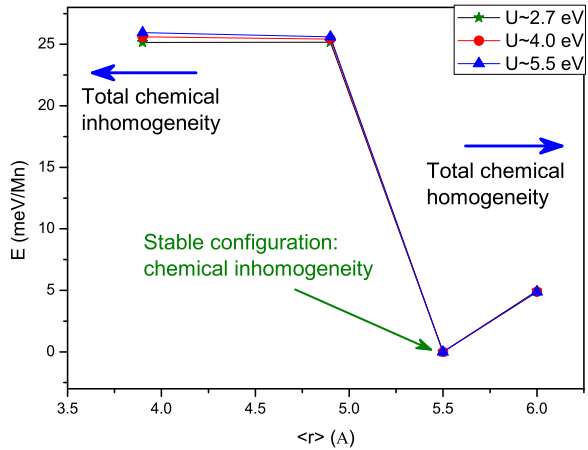}
\caption{(Color online) Analogous study of energies for La$_{0.625}$Ca$_{0.375}$MnO$_{3}$ than Fig. \ref{la08_r}. Calculations were made in a FM configuration (the more stable at this concentration) at several values of U, leading to similar results than La$_{0.8125}$Ca$_{0.1875}$MnO$_{3}$. }\label{la06_r}
\end{figure}

In order to further check the reliability of these calculations, we have analyzed a different Ca$^{2+}$ concentration ($x = 0.375$) which is not at the boundary between different ground states in the La$_{1-x}$Ca$_{x}$MnO$_{3}$ solid solution. For this, we have constructed four different possible Ca-dopant distributions in the La$_{0.625}$Ca$_{0.375}$MnO$_{3}$.\cite{yang,pineiro_pssc} A similar energy analysis as the one performed above shows that the chemical inhomogeneities that occur at this concentration are even larger in magnitude. The displacements in the local concentration of the $<r>$= 5.5 \AA\ structure, that can be observed in Fig. \ref{la06_r}, ranges from $0.335<x<0.407$. Note that $<r>$ is smaller than in the $x = 0.1875$ case (i.e. 6.71 \AA\ ). This is related to the richer calcium concentration in $x = 0.375$. The above range of Ca$^{2+}$ concentrations is always far enough from the FM insulator ground state at $x = 0.18$ as well as the AFM with CO ground state observed at $x = 0.5$. In spite of this, experimental evidence of inhomogeneities in manganites has been obtained.\cite{pr_magnanites} Thus, chemical inhomogeneities are always present in this kind of samples, but a proximity to a phase boundary is required for them to act as nucleation centers of an electronic/magnetic phase separation scenario.

\section{Conclusions}

Our results show important evidences about an often forgotten factor (mainly due to the complications for its direct calculation and measurement) in studying the origin of electronic and magnetic phase separation close to a magnetic phase transition. By means of ab initio calculations, we have studied mixed magnetic phases but none of them is stable. However, chemical inhomogeneities caused by cation disorder in a scale smaller than 1 nm are stable in the system according to our total energy calculations. Close to a FM/AF phase transition, these nanoscale doping inhomogeneities are large enough to produce hole-rich/hole-poor regions in the compound at the nanometer scale. For some concentrations, like the $x$ $\sim$ 0.2 phase boundary, they are found to have a similar size than the observed phase separated regions. Thus, their effects cannot be ignored and could be substantial, whether driving electronic/magnetic phase separation or accompanying it. Those cationic inhomogeneities could act as charge attractors and/or nucleation centers for the phenomenon of phase separation to develop. If the scale of the phase separation phenomenon were micrometric, or much larger than 1 nm, these chemical effects could possibly be ignored.

\section{Acknowledgments}

We thank F. Rivadulla and P. Blaha for fruitful discussions. We also thank the CESGA (Centro de Supercomputacion de Galicia) for the computing facilities and the Ministerio de Educaci\'on y Ciencia (MEC) for the financial support through the projects MAT2006/10027 and MAT2009-08165. We are also thankful to the Xunta de Galicia for financial support through the project INCITE 08PXIB236052PR.

%\bibliography{phase_separation}

\end{document}